\documentclass{aastex63}
\submitjournal{ApJ}
\newcommand{\MyFigA}{\ref{MyFigA}}
\newcommand{\MyFigB}{\ref{MyFigB}}
\newcommand{\MyFigC}{\ref{MyFigC}}
\newcommand{\MyFigD}{\ref{MyFigD}}

\newcommand{\MyTabA}{\ref{MyTabA}}
\newcommand{\R}{\textcolor[rgb]{1.00,0.00,0.00}}

\newcommand{\Bezier}{B\'{e}zier }

\begin{document}
\title{Electron Spectrum for the Prompt Emission of Gamma-ray Bursts in the Synchrotron Radiation Scenario}
\correspondingauthor{Da-Bin Lin}
\email{lindabin@gxu.edu.cn}
\author{Kuan Liu}
\author{Da-Bin Lin}
\author{Jing Li}
\author{Yu-Fei Li}
\author{Rui-Jing Lu}
\author{En-Wei Liang}
\affiliation{GuangXi Key Laboratory for Relativistic Astrophysics, School of Physical Science and Technology, Guangxi University, Nanning 530004, China}
\begin{abstract}
Growing evidences indicate that the synchrotron radiation mechanism
may be responsible for the prompt emission of gamma-ray bursts (GRBs).
In the synchrotron radiation scenario,
the electron energy spectrum of the prompt emission is diverse in theoretical works
and has not been estimated from observations in a general way
(i.e., without specifying a certain physical model for the electron spectrum).
In this paper, we creatively propose a method to directly estimate
the electron spectrum for the prompt emission,
without specifying a certain physical model for the electron spectrum {\bf in the synchrotron radiation scenario}.
In this method, an empirical function (i.e., a four-order \Bezier curve jointed with a linear function at high-energy) is applied to describe the electron spectrum in log-log coordinate.
It is found that our empirical function can well mimic the electron spectra obtained in many numerical calculations or simulations.
Then, our method can figure out the electron spectrum for the prompt emission without specifying a model.
By employing our method on observations, taking GRB~180720B and GRB~160509A as examples, it is found that the obtained electron spectra are generally different from that in the standard fast-cooling scenario and even a broken power law.
Moreover, the morphology of electron spectra in its low-energy regime varies with time in a burst and even in a pulse. Our proposed method provides a valuable way to confront the synchrotron radiation mechanism with observations.
\end{abstract}


\section{Introduction}\label{Sec:Introduction}

The radiation mechanism for the prompt emission of gamma-ray bursts (GRBs) remains unclear after decades of observations.
The radiation spectra of the prompt emission are usually characterized by an exponential-jointed broken power-law function, i.e., Band function (\citealp{Band_D-1993-Matteson_J-ApJ.413.281B}).
The typical value of parameters in Band function by fitting the observations are $\alpha\sim-1$, $\beta\sim-2.2$, and $E_{\rm p}\sim400$keV, where $\alpha$, $\beta$, and $E_{\rm p}$ are the low-energy photon spectral index, high-energy photon spectral index, and the peak photon energy, respectively (\citealp{Preece_RD-2000-Briggs_MS-ApJS.126.19P}; \citealp{Nava_L-2011-Ghirlanda_G-A&A.530A.21N}; \citealp{Kaneko_Y-2006-Preece_RD-ApJS.166.298K}; \citealp{Goldstein_A-2012-Burgess_JM-ApJS.199.19G}).
Owing to the lack of physical origin for Band function,
one derives the physical implications by inferring what mechanism the fit parameters can be produced by.
Synchrotron radiation is a very natural candidate to explain the non-thermal feature of Band function (\citealp{Meszaros_P-1994-Rees_MJ-ApJ.432.181M}; \citealp{Tavani_M-1996-ApJ.466.768T}; \citealp{Daigne_F-1998-Mochkovitch_R-MNRAS.296.275D}; \citealp{Ghirlanda_G-2002-Celotti_A-A&A.393.409G}).
However, the most straightforward synchrotron model suffers from ``fast-cooling problem'',
i.e., the typical observed spectrum should have a low-energy photon spectral index  $-3/2$, which strongly conflicts with observations (\citealp{Sari_R-1998-Piran_T-ApJ.497L.17S}; \citealp{Ghisellini_G-2000-Celotti_A-MNRAS.313L.1G}).
Many attempts have been made to alleviate the fast-cooling problem, e.g., adopting a decaying magnetic field in emission region (\citealp{Peer_A-2006-Zhang_B-ApJ.653.454P}; \citealp{Uhm_ZL-2014-Zhang_B-NatPh.10.351U}; \citealp{Zhao_XH-2014-Li_Z-ApJ...780...12Z}), introducing a slow heating mechanism by magnetic turbulence (\citealp{Asano_K-2009-Terasawa_T-ApJ...705.1714A}), involving the inverse Compton scattering effect at the Klein-Nishina regime (\citealp{Derishev_EV-2001-Kocharovsky_VV-A&A.372.1071D}; \citealp{Nakar_E-2009-Ando_S-ApJ...703..675N}), considering a marginally fast cooling regime (\citealp{Daigne_F-2011-Bosnjak_Z-A&A.526A.110D}; \citealp{Beniamini_P-2018-Barniol_DR-MNRAS.476.1785B}; \citealp{Florou_l-2021-Petropoulou_M-arXiv210202501F}),
or invoking a fast-increasing electron energy injection rate (\citealp{Liu_K-2020-Lin_DB-ApJ.893L.14L}).
In addition, it is shown that the synchrotron model could not account for about one third of bursts with $\alpha>-2/3$, which is the so-called synchrotron ``line-of-death'' problem (\citealp{Preece_RD-1998-Briggs_MS-ApJ.506L.23P}).
There have been numerous studies proposed to break the line-of-death limit, such as,
considering the synchrotron self-absorption (\citealp{Preece_RD-1998-Briggs_MS-ApJ.506L.23P}),
jitter radiation within small-scale random magnetic field (\citealp{Medvedev_MV-2000-ApJ.540.704M}; \citealp{Mao-JR-2013-Wang_JC-ApJ.776.17M}),
the synchrotron emission from the relativistic electrons with a small pitch angle (\citealp{Lloyd_NM-2000-Petrosian_V-ApJ.543.722L}; \citealp{Lloyd_NM-2002-Petrosian_V-ApJ.565.182L}; \citealp{Yang_YP-2018-Zhang_B-ApJ.864L.16Y}),
or involving the inverse Compton scattering effect (\citealp{Liang_E-1997-Kusunose_M-ApJ.479L.35L}).

On the other hand, it has been proven that directly fitting the observations based on the
synchrotron radiation model can be also effective.
By adopting electron spectra being composed of a thermal Maxwell distribution connected to a power law
at high-energy,
\cite{Tavani-1996-ApJ.466.768T}, \cite{Baring-2004-Braby-ApJ.613.460B}, and \cite{Burgess_JM-2014-Preece_RD-ApJ.784.17B}
fit the observed radiation spectrum in the synchrotron radiation scenario with or without a photospheric emission.
\cite{Lloyd_NM-2000-Petrosian_V-ApJ.543.722L} and \cite{Lloyd_NM-2002-Petrosian_V-ApJ.565.182L}
investigate the synchrotron emission models as the source of GRB prompt emission spectra
by involving the ``smooth cutoff'' to the electron spectrum.
To test the radiation mechanism of the prompt emission,
\cite{Oganesyan_G-2019-Nava_L-A&A.628A.59O} adopted broken power law electron spectra in synchrotron model to fit the prompt emission with optical observations.
\cite{Zhang_BB-2016-Uhm_ZL-ApJ.816.72Z} and \cite{Burgess_JM-2020-Damien-NatAs.4.174B} fit the observations
in the synchrotron radiation scenario by specifying a physical model for the electron spectra.
Although many efforts have been done,
the functional forms adopted to describe the electron spectra are generally model-dependent.
It is also worth to point out that the electron spectrum can be very diverse in numerical calculations or simulations (e.g., \citealp{Uhm_ZL-2014-Zhang_B-NatPh.10.351U}, \citealp{Sironi_L-2009-Spitkovsky_A-ApJ.698.1523S}, \citealp{Guo_F-2014-Li_H-PhRvL.113o5005G}, and \citealp{Liu_K-2020-Lin_DB-ApJ.893L.14L}).
Please see Figure~{\MyFigA} for a glance of some examples (dashed lines).
In these cases, using a model-dependent electron spectrum in the synchrotron radiation scenario
to fit the observations may bias the understand of the prompt emission.
{\bf In this paper,
we propose a method in this paper to directly estimate the electron spectrum
for the prompt emission,
without specifying a certain physical model or presumptive morphology for the electron spectrum.
}
This paper is organized as follows.
In Section~\ref{Sec:model}, we describe our proposed empirical function in details.
The empirical function is the key point of our method and
{\bf we focus on the synchrotron radiation scenario}.
In Section~\ref{Sec:SpecAnalysis}, our method is applied to discuss Band radiation spectrum and on spectral analysis of observations.
In Section~\ref{Sec:Conclusion}, we summary our results.

\section{Prescription of electron spectrum and Fitting Method}\label{Sec:model}
{\bf In this paper, we mainly focus on how to estimate the electron spectrum
for the prompt emission in the synchrotron radiation scenario.}
In the synchrotron radiation scenario, the GRBs' prompt emission is generated from a group of relativistic electrons.
Therefore, the electron energy spectrum for the prompt emission can be estimated by fitting the radiation spectrum.
For this purpose, we propose an empirical function to picture out the possible electron spectrum.

\begin{figure}
\begin{center}
\begin{tabular}{ccc}
\includegraphics[angle=0,scale=0.3,trim=200 100 60 0]{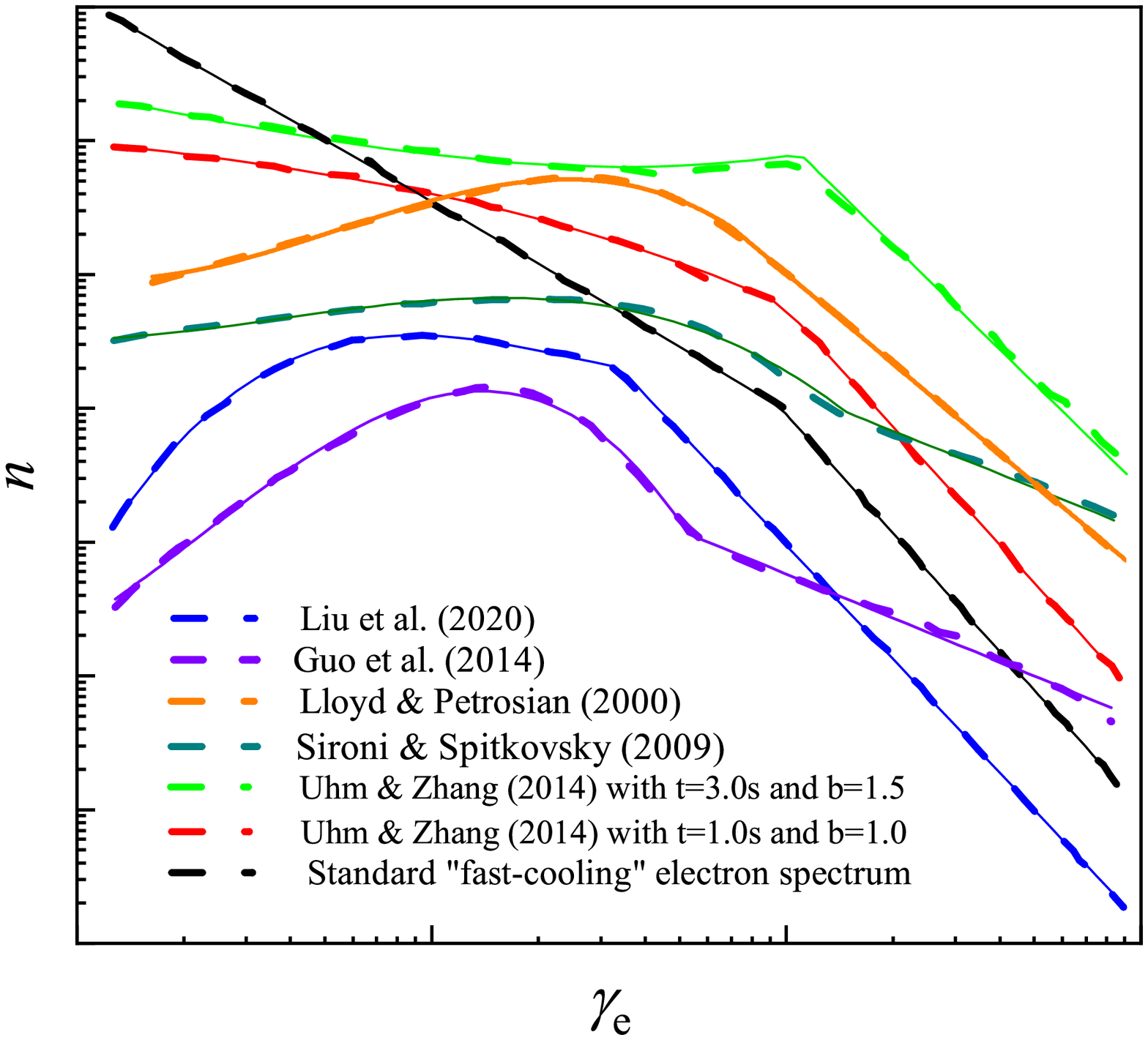} &
\includegraphics[angle=0,scale=0.245,trim=80 20 0 0]{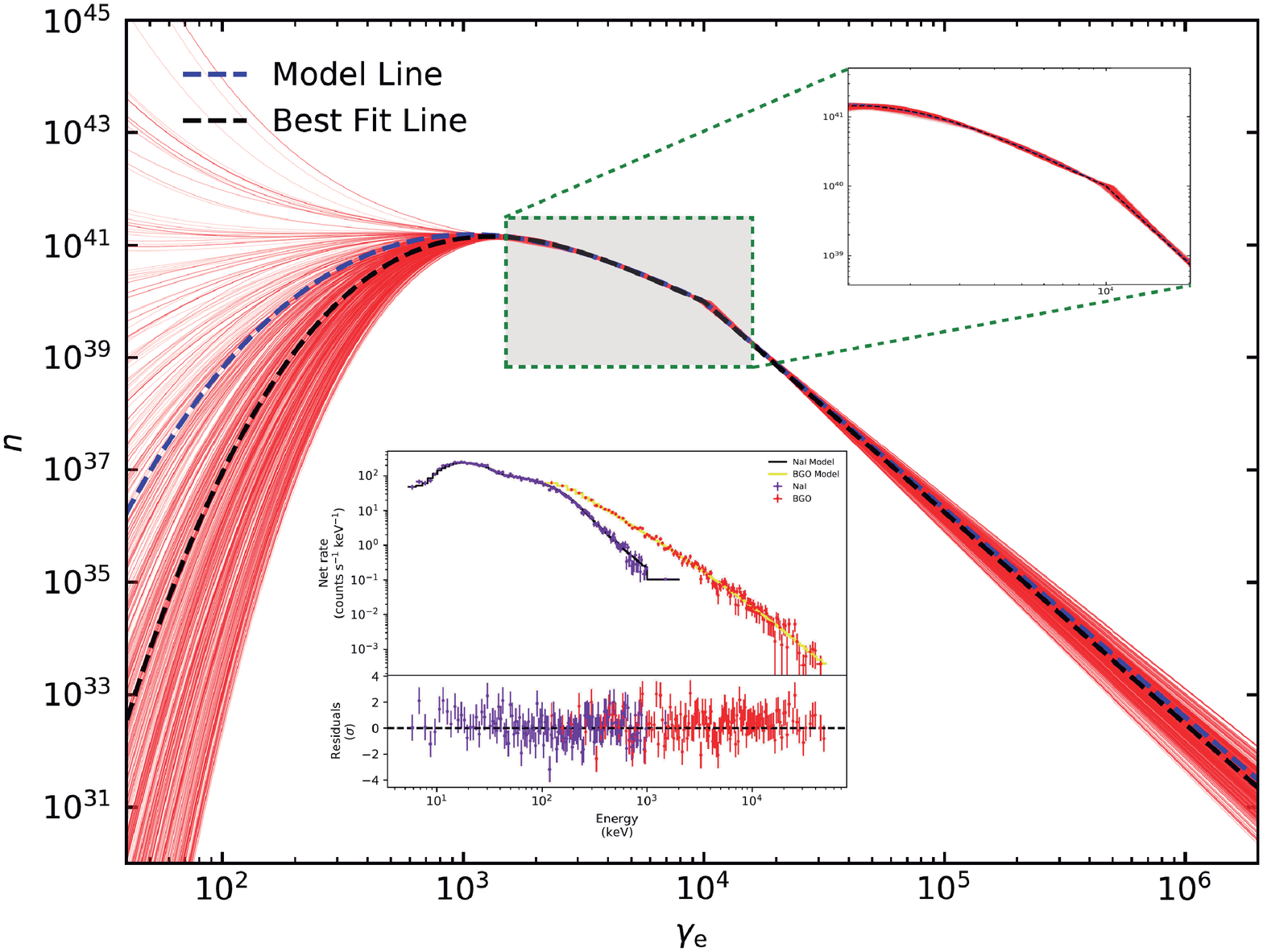} &
\includegraphics[angle=0,scale=0.18,trim=30 50 0 0]{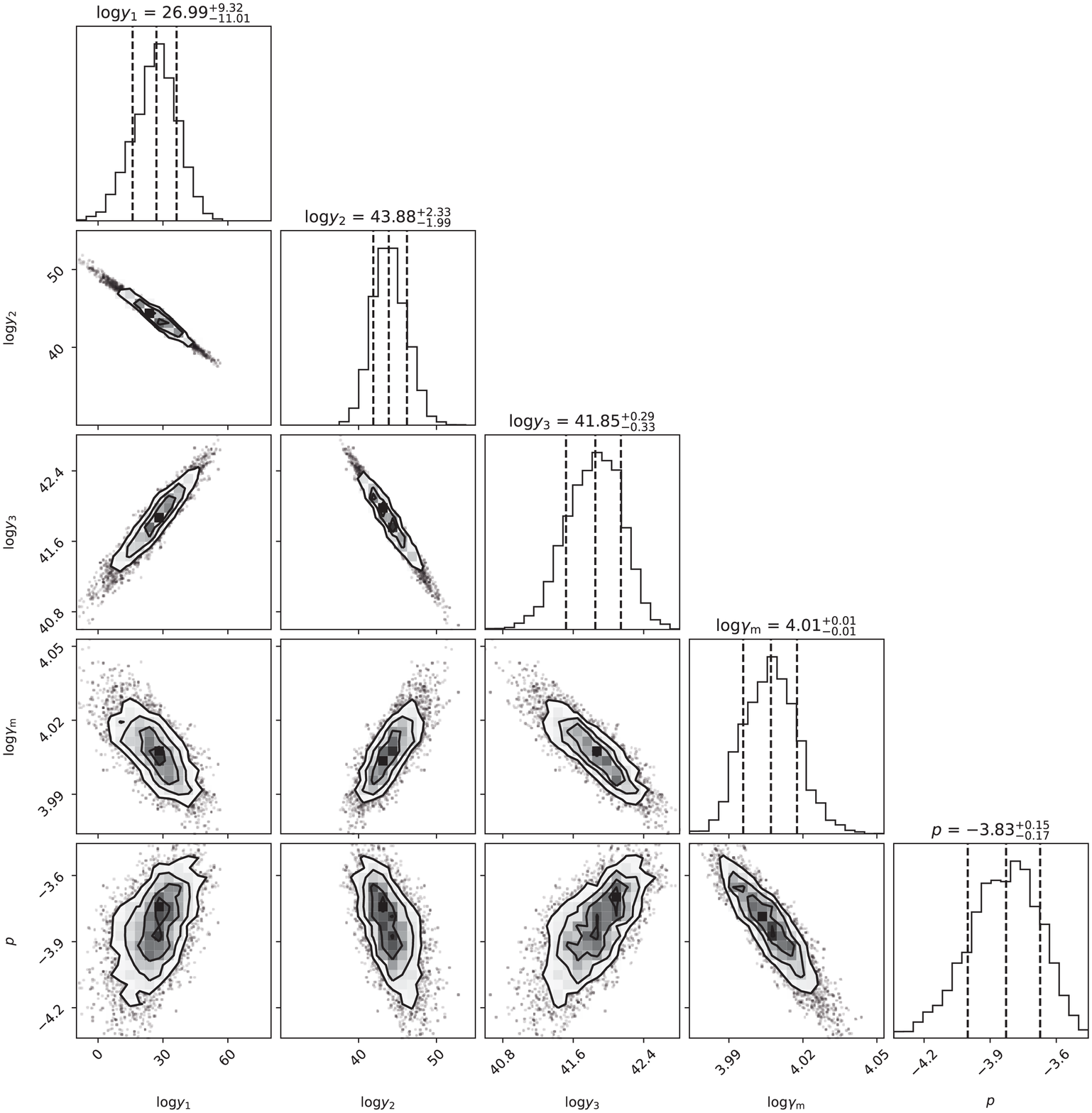}
\end{tabular}
\end{center}
\caption{Testing of our empirical function (left panel) and the spectral-fitting-method (middle and right panel).
\emph{Left-panel}, the electron spectra collected from different works and the corresponding best fitting results with our empirical function are shown with dashed and solid lines, respectively.
Here, the purple, dark green, green, red, orange, and blue dashed lines are the electron spectra obtained from
the figure~1 of \cite{Guo_F-2014-Li_H-PhRvL.113o5005G} with $\omega_{\rm pe}t$=700,
the figure~10 of \cite{Sironi_L-2009-Spitkovsky_A-ApJ.698.1523S} with $\theta=30^\circ$,
the second panel of the figure~1 in \cite{Uhm_ZL-2014-Zhang_B-NatPh.10.351U} with $t_{\rm obs}=1.0$s,
the forth panel of the figure~1 in \cite{Uhm_ZL-2014-Zhang_B-NatPh.10.351U} with $t_{\rm obs}=3.0$s,
the equation~4 of \cite{Lloyd_NM-2000-Petrosian_V-ApJ.543.722L} with $q=1.0$ and $p=3.0$,
and the figure~1 of \cite{Liu_K-2020-Lin_DB-ApJ.893L.14L} with $t_{\rm obs}=1.2$s,
respectively.
{\bf
\emph{Middle-panel},
the electron spectra based on the last $1\%$ iterations from MCMC sampling
are plotted with red lines,
where the blue and black dashed line
represent the given electron spectrum and the best fitting result for the electron spectrum from MCMC sampling.
In addition, the upper inset shows zoomed-in view for the electron spectrum at $\gamma_{\rm e}\sim 10^3-10^4$
and the bottom inset shows the best fitting result on the synthetic data.}
\emph{Right-panel},
the posterior probability density functions by applying
our spectral-fitting-method on the synthetic data.}\label{MyFigA}
\end{figure}

Intuitively, the electron spectrum shown in Figure~{\MyFigA} can be decomposed into two segments: a high-energy segment and a low-energy segment jointed at the electron Lorentz factor $\gamma_{\rm e}=\gamma_{\rm m}$.
The high-energy segment usually relates to the electron injection rate $Q(\gamma_{\rm e})\propto\gamma_{\rm e}^{-p}$ and can be described by a power-law function $n(\gamma_e)\propto\gamma_e^{-p-1}$,
where $nd\gamma_e$ is the number of electrons in $[\gamma_{\rm e}, \gamma_{\rm e}+d\gamma_{\rm e}]$.
However, the morphology of the low-energy segment is diversity in theoretically and can be very different from a power-law function.
Then, we introduce a four-order \Bezier curve
\footnote{
\Bezier curve is a smooth curve defined by some given control points,
which is wildly used in computer graphics and the related fields.
In this paper, we adopt a simple four-order, two-dimensional \Bezier curve,
which is created by four control points $P_1$, $P_2$, $P_3$, and $P_4$ in two-dimensional space.
In general, it starts at $P_1$ going toward $P_2$ and arrives at
$P_4$ coming from the direction of $P_3$.
Usually, it would not pass through $P_2$ and $P_3$ unless these four
points are in a line.
However, these two points would determine the behavior
of \Bezier curve between $P_1$ and $P_4$.
}
to describe the low-energy electron spectrum in log-log coordinate (i.e., the $\log \gamma_e-\log n$ plane).
Therefore, our empirical function used to describe the electron spectrum is read as
\begin{equation}\label{Eq:Electrons_Spectrum}
\log {n} = \left\{ {\begin{array}{*{20}{c}}
{{\mathbb{B}}({\gamma_{\rm e}}),}&{{\gamma_{\rm e}} < {\gamma_{\rm{m}}},}\\
{\log [{y_{\rm m}}{{\left( {{\gamma_{\rm e}}/{\gamma _{\rm{m}}}} \right)}^{ -p-1}}],}&{{\gamma_{\rm e}} \ge {\gamma_{\rm{m}}},}
\end{array}} \right.
\end{equation}
where $\mathbb{B}(\gamma_{\rm e})$ is the four-order \Bezier curve and
$y_{\rm m}$ is the number density of electrons at $\gamma_{\rm e}=\gamma_{\rm m}$.
The four-order \Bezier curve $\mathbb{B}(\gamma_{\rm e})$ is described with a serial of points $(\gamma_{\rm e}(t), \mathbb{B}(t))$, which are calculated with following equation
by varying $t$ from $0$ to $1$,
\begin{equation}\label{Eq:Bezier}
\left\{ \begin{array}{l}
\mathbb{B}(t)= {(1 - t)^3}\log {y_1} + 3t{(1 - t)^2}\log {y_2} + 3{t^2}(1 - t)\log {y_3} + {t^3}\log {y_{\rm m}},\\
\log\gamma_{\rm e}(t) = {(1 - t)^3}\log {\gamma_{\rm e,1}} + 3t{(1 - t)^2}\log {\gamma_{\rm e,2}} + 3{t^2}(1 - t)\log {\gamma_{\rm e,3}} + {t^3}\log {\gamma_{\rm m}}.
\end{array} \right.
\end{equation}
Here, $P_1(\log {\gamma_{\rm e,1}}, \log {y_1})$, $P_2(\log {\gamma_{\rm e,2}}, \log {y_2})$,
$P_3(\log {\gamma_{\rm e,3}}, \log {y_3})$, and $P_4(\log {\gamma_{\rm m}}, \log {y_{\rm m}})$
are four control points used to create the \Bezier curve.
To simply our fittings, we adopt $\log\gamma_{\rm e,1}=1$, $\log\gamma_{\rm e,2}=({\rm log} \gamma_{\rm m,0} -{\rm log} {\gamma_{\rm e,1}})/3+{\rm log} \gamma_{\rm e,1}$, and $\log\gamma_{\rm e,3}=2({\rm log} \gamma_{\rm m,0} -{\rm log} {\gamma_{\rm e,1}})/3+{\rm log} \gamma_{\rm e,1}$, where $\gamma_{\rm m,0}=10^4$ is set as the initial value of $\gamma_{\rm m}$.
Then, the free parameters in our empirical function are $y_1$, $y_2$, $y_3$, $\gamma_{\rm m}$, and $p$.
We fit the electron spectra in the left panel of Figure~\ref{MyFigA} with Equation~(\ref{Eq:Bezier}), where the fitting results are shown with solid lines in this panel.
One can find that the electron spectra in the left panel of Figure~\ref{MyFigA} can be well described with our empirical function.
Then, our empirical function can be used to figure out the electron spectrum for the prompt emission,
without specifying a certain physical model for the electron spectrum.

It should be noted that the electron spectrum, which can be described with our empirical function,
should be continuous.
If not, such as the electron spectrum in the figure 3 of \cite{Burgess_JM-2011-Preece_RD-ApJ.741.24B},
our empirical function could not present a well fit.
{\bf In addition, freeing the electron spectrum is not equivalent to having an empirical photon spectrum in the first place.
Firstly, the lowest power-law index of the photon spectrum from our empirical electron spectrum in the synchrotron radiation scenario should be larger than $-2/3$.
Secondly, the electron spectrum for prompt emission carry the information from the particle accelerating and cooling mechanism.
Thus the estimation for electron spectrum could help us to better understand the energy dissipation process in relativistic jet.
}

For a given electron spectrum, the observed synchrotron radiation flux at a given frequency $\nu$ can be calculated as
\begin{equation}\label{EQ:SynPower}
{f_\nu }(\nu)= \frac{{\sqrt 3 q_{\rm{e}}^3B\Gamma }}{{2\pi d_L^2{m_{\rm{e}}}{c^2}}}\int_{\gamma_{\rm e,1}}^\infty  F \left( {\nu /{\nu _{\rm{c}}}} \right){n}({\gamma _e})d{\gamma _e},
\end{equation}
where $F(x)=x\int_{x}^{+\infty}K_{5/3}(k)dk$, $K_{5/3}(k)$ is the modified Bessel function of 5/3 order,
$\nu_{\rm c}=3q_{\rm e}B^2\gamma_{\rm e}^2\Gamma(1+z)/(2\pi m_{\rm e}c)$ is the characteristic frequency of the electron with Lorentz factor $\gamma_{\rm e}$ in magnetic field $B$,
$\Gamma=300$ is the bulk Lorentz factor of the jet,
$d_{\rm L}$ is the luminosity distance,
and $q_{\rm e}$, $m_{\rm e}$, and $c$ are the electron charge, electron mass, and light speed, respectively.

Based on the Equations~(\ref{Eq:Electrons_Spectrum}) and (\ref{EQ:SynPower}),
we can fit the radiation spectrum of the prompt emission and obtain the corresponding electron spectrum.
This is our proposed spectral-fitting-method used to estimate the electron spectrum
for the prompt emission in the synchrotron radiation scenario.
To test our method, we perform a simple testing fitting on a synthetic data.
{\bf
The synthetic data is generated as follows:
\textcircled{\footnotesize{i}} We create a synchrotron radiation spectrum based on a bump-shape electron spectrum.
As an example, the black-dashed line in the middle panel of Figure~\ref{MyFigA},
i.e., Equation~(\ref{Eq:Electrons_Spectrum}) with $\log{\gamma_{\rm e,1}}=1$, $\log{\gamma_{\rm m}}=4$,
$\log{y_1}=30$, $\log{y_2}=43$, $\log{y_3}=42$, $\log{y_{\rm m}}=40$, and $p=-3.7$,
is adopted as our electron spectrum. In addition, $B=30$~Gs is took.
\textcircled{\footnotesize{ii}} We fold this synchrotron radiation spectrum
through the instrumental response of the Fermi Gamma-ray Burst Monitor
to create a poisson-distributed synthetic data,
where the python source package {\tt threeML}
\footnote{\url{https://github.com/threeML/threeML}} (\citealp{Vianello_G-2015-arXiv}) is used.}
Then, we perform the spectral fitting based on the synthetic data.
The spectral fitting is performed based on
the Markov Chain Monte Carlo (MCMC) method
to produce posterior predictions for the model
parameters, i.e., $\log{y_1}$, $\log{y_2}$, $\log{y_3}$, $\log{\gamma_{\rm m}}$, and $p$.
The python source package {\tt emcee}
\footnote{\url{https://github.com/dfm/emcee/blob/b9d6e3e7b1926009baa5bf422ae738d1b06a848a/docs/index.rst}}
(\citealp{Foreman-2013-Hogg-PASP.125.306F})
is used for our MCMC sampling,
{\bf where $N_{\rm walkers}\times N_{\rm steps}=10\times 10^5$ is adopted
and the initial $50\%$ iterations are used for burn-in.
The priors of $\log{y_1}$, $\log{y_2}$, $\log{y_3}$, $\log{\gamma_{\rm m}}$, and $p$
are set as uniform distribution in the range of (-30, 100), (10, 70), (30, 50), (3, 5), and (-5, -3),
respectively.\footnote{
The priors of $\log{y_1}$, $\log{y_2}$, and $\log{y_3}$
are set based on the following consideration.
With Equation~(\ref{Eq:Bezier}),
we fit the electron spectra in the left panel of Figure~\ref{MyFigA}.
The fitting results reveal that the values of $\log{y_3}$ and $\log{y_2}$ do not
deviate from the value of $\log{y_4}$ significantly.
Therefore, we set the priors of $\log{y_3}$ and $\log{y_2}$ as $(\log{y_4}-10,\,\log{y_4}+10)$
and $(\log{y_4}-30,\,\log{y_4}+30)$, i.e., (30, 50) and (10, 70), respectively.
In addition, the prior of $\log{y_1}$ may be in a wide range.
The reason can be found in the end of Section~\ref{Sec:model}.
Then, we set the prior of $\log{y_1}$ as $(-30,\,100)$.
Actually, we also try a wider range of the priors for these three parameters
and obtain very similar fit results.}}
{\bf The projections of the posterior distribution in 1D and 2D for the model parameters
are presented in the right panel of Figure~\ref{MyFigA}
and the electron spectra based on the last $1\%$ iterations
are also plotted in the middle panel of Figure~\ref{MyFigA} with red lines.
One can find that the obtained values of $\log{y_3}=41.85_{-0.33}^{+0.29}$, $\log{\gamma_{\rm m}}=4.01_{-0.01}^{+0.01}$, and $p=-3.83_{-0.17}^{+0.15}$
are similar to those of our provided electron spectrum.
However, the obtained values of $\log{y_1}=26.99_{-11.01}^{+9.32}$ and $\log{y_2}=43.88_{-1.99}^{+2.33}$ deviate from those of our
provided electron spectrum, especially for the value of $\log{y_1}$.
It implies that the electron spectrum from our spectral fittings
are only robust in the low-energy and high-energy ranges rather than the lowest-energy range.}

\section{Spectral Analysis}\label{Sec:SpecAnalysis}

\subsection{Comments on Band Function}
In this subsection, we investigate the electron spectrum related to Band radiation spectrum in synchrotron radiation scenario.
A Band function with typical parameters $\alpha=-1$, $\beta=-2.3$, and $E_{\rm p}=400$ keV is discussed in this subsection and shown in Figure~{\MyFigB} with black line.
In general, it is believed that such radiation spectrum is originated from the synchrotron radiation of a broken power-law electron spectrum with $p_{\rm low}=2(\alpha+1)-1$ and $p=2(\beta+1)-1$,
where $p_{\rm low}$ and $p$ are the low-energy and high-energy power-law indexes, respectively.
The synchrotron radiation spectrum of such electron spectrum is shown in Figure~{\MyFigB} with green dashed line.
Obviously, the radiation
spectrum generated from such kind of broken power-law electron spectrum is very different from Band radiation
spectrum, especially for the part around the transition from low-energy spectral segment to high-energy spectral
segment.
The transition is apparently sharp for Band function compared with the synchrotron radiation spectrum.
This behavior has also been found in many previous works,
e.g., \cite{Zhang_BB-2016-Uhm_ZL-ApJ.816.72Z} and \cite{Burgess_JM-2019-A&A.629A.69B}.
This result suggests that the Band radiation spectrum may not be produced by a broken power-law electron spectrum.

In the following, we search for the most suitable electron spectrum for Band radiation spectrum by fitting it with Equations~(\ref{Eq:Electrons_Spectrum}) and (\ref{EQ:SynPower}).
The obtained electron spectrum and its radiation spectrum are shown in Figure~{\MyFigB} with red dashed line.
Although the obtained radiation spectrum is basically consistent with Band radiation spectrum, it is a bit weird for the unexpected sharp peak at $\gamma_{\rm m}$ and the strange bump at the low-energy regime of electron spectrum.
We point out that this kind of electron spectrum may be unnatural to some degree.
The reasons are shown as follows.
(1) The peak at $\gamma_{\rm m}$ is mainly related to the exponential-connected break in Band function, whereas the physical origin of this break is no clear yet.
(2) Although the obtained electron spectrum can produce a Band-like synchrotron radiation spectrum, the position of low-energy bump and $\gamma_{\rm m}$-peak in electron spectrum should be fine-tuned, which may hardly exist in real situation.
(3) The shape of this electron spectrum is very different from those in the left panel of Figure~{\MyFigA}, except the one shown with green line, which has a similar peak at $\sim \gamma_{\rm m}$.
However, one should note that such kind of electron spectrum mainly appears without making significant contribution to the observed flux (e.g., \citealp{Uhm_ZL-2014-Zhang_B-NatPh.10.351U}).
Therefore, we would like to believe that the $\gamma_{\rm m}$-peak in the electron spectrum corresponding to Band function may be an unnatural outcome.
Then, the exponential transition in Band function may not well describe the transition behavior in the radiation spectrum of the prompt emission if the synchrotron radiation does work.

This subsection is dedicated to study the electron spectrum corresponding to Band radiation spectrum in the synchrotron radiation scenario.
We found that the electron spectrum of the Band radiation spectrum may be hardly
reproduced in a physical model, e.g., the models producing the electron spectrum in Figure~{\MyFigA}.
It suggests that the Band radiation spectrum may be not intrinsic to the prompt emission of GRBs,
especially to the transition segment (between low-energy regime and high-energy regime) in the radiation spectrum.
We would like to point out that to understand the characteristics of the Band radiation spectrum,
fitting the synthetic observed data of the synchrotron radiation with the Band function is necessary.
For example, \cite{Burgess_JM-2015-Ryde_F-MNRAS.451.1511B} simulate synchrotron or synchrotron+blackbody spectra and fold them through the instrumental response of the Fermi Gamma-ray Burst Monitor.
They then perform a standard data analysis by fitting the synthetic data with both Band and Band+blackbody models to investigate the ability of the Band function to fit a synchrotron spectrum within the observed energy band.

\subsection{Application on GRBs~180720B and 160905A} \label{SubSec:Application}

In this subsection, we fit the radiation spectrum of GRBs~180720B and 160905A with Equations~(\ref{Eq:Electrons_Spectrum}) and (\ref{EQ:SynPower})
to estimate the electron spectrum in the synchrotron radiation scenario.
In our spectral analysis, we use the data from the \emph{Fermi}/GBM.
GBM has 12 sodium iodide (NaI) scintillation detectors covering the 8~keV-1~MeV energy band, and two bismuth germanate (BGO) scintillation detectors being sensitive to the 200~keV-40~MeV energy band (\citealp{Meegan_C-2009-Lichti_G-ApJ.702.791M}).
The brightest NaI and BGO detectors are used in our analyses.
The python source package {\tt gtBurst}\footnote{\url{https://github.com/giacomov/gtburst}} is used to extract the light curves and source spectra.
 {\tt Xspec} (\citealp{Arnaud-1996-ASPC.101.17A}; \citealp{Atwood_WB-2009-Abdo_AA-ApJ.697.1071A}) is used to perform spectral analysis\footnote{
The initial values of $y_1$, $y_2$, $y_3$, and $p$ are set as follows.
Firstly, the prompt emission is fitted with Band function to obtain the optimum value of $\alpha$, $\beta$, and $E_0$.
Then, $B$ can be set by solving $\nu_{\rm b}\equiv 0.3\times3q_{\rm e}B^2\gamma_{\rm m,0}^2\Gamma(1+z)/(2\pi m_{\rm e}c)=E_{\rm b}\equiv E_0(\alpha-\beta)$,
where $E_{\rm b}$ is the break photon energy of Band function.
In addition, the electron spectrum is initially set as a broken power law with $p_{\rm low}=(\alpha+1)\times2-1$
and $p=(\beta+1)\times2-1$.
$y_{\rm m}$ is set by equaling $f_{\nu}(\nu_{\rm b})/\nu_{\rm b}$ to the photon flux of the Band function at $E_{\rm b}$. In our fitting, $y_1$, $y_2$, $y_3$, $\gamma_{\rm m}$ and $p$ are the free parameters.
Based on the above settings,
we perform a tentative spectral fit to roughly estimate parameters in a relatively wide parameter areas.
With the obtained optimum fitting results from the tentative fitting,
we further perform a fine spectral fitting based on a narrow parameter areas.
},
where the ``Poisson-Gauss'' fit statistic (i.e., pgstat) is adopted.
The theoretical electron spectra from numerical calculations or simulations are almost a bump or power-law shape
in its low-energy regime (see the left panel of Figure~{\MyFigA}).
Then, Equation~(\ref{Eq:Electrons_Spectrum}) is restricted to be a bump or power-law shape in our fittings.
That is to say, the point $P_2$ ($P_3$) should be above or on the line of $P_1P_4$ ($P_2P_4$).

{\bf \emph{GRB~180720B Analysis }}
GRB~180720B is a long burst with a redshift $z=0.654$ and detected by \emph{Fermi} and \emph{Swift} satellites (\citealp{Roberts_OJ-2018-Meegan_C-GCN.22981....1R}, \citealp{Bissaldi_E-2018-Racusin_JL-GCN.22980....1B}, \citealp{Siegel_MH-2018-GCN.22973....1S}, \citealp{Vreeswijk_PM-2018-Kann_DA-GCN.22996....1V}).
The obtained NaI 6 light curve of the prompt emission is shown in the bottom inset of each panel in Figure~{\MyFigC},
where the brightest NaI (i.e., NaI 6 and NaI 8) and BGO (i.e., BGO 0) detectors are used in our analyses.
As an example, we first select a time period of $[7.14,\,8.19]$~s after the burst triggered for our analysis, which is marked with blue color in the bottom inset of the left panel in Figure~{\MyFigC}.
This time period is also used in the spectral analysis of \cite{Ravasio_ME-2019-Ghirlanda_G-A&A...625A..60R}, of which the results can be used to compare with ours.
The spectral fitting result is shown with black line in the upper inset of the left panel.
The corresponding electron spectrum is shown with blue solid line in this panel
and also reported in Table~\ref{MyTabA}.
{\bf Inspired by the fit result in Section~\ref{Sec:model},}
such kind of electron spectrum can be decomposed into three segments: the lowest-energy segment (marked with pink shadow), the low-energy segment (marked with yellow shadow), and the high-energy segment (marked with cyan shadow).
{\bf It should be note that only the low-energy segment and the high-energy segment are robust in our analysis.
The reason is presented at the end of this section.
}
One can find that the low-energy segment at $\gamma_{\rm e}\sim \gamma_{\rm m}$ can be approximated as $n_{\rm e}\propto \gamma_{\rm e}^{-2}$,
which is the low-energy electron spectrum in the standard fast-cooling pattern and is shown with black dashed line in Figure~{\MyFigC}.
This result is consistent with what reported in \cite{Ravasio_ME-2019-Ghirlanda_G-A&A...625A..60R}.
Therefore, our method is applicable to estimate the electron spectrum for the prompt emission in the synchrotron radiation scenario.

For the pulse in [7.14, 9.00]~s,
we also perform detail spectral analysis on the remaining time periods,
e.g., [8.19, 8.70]~s and [8.70, 9.00]~s,
which are marked with red and green colors in the inset of middle panel of Figure~{\MyFigC}, respectively.
The obtained electron spectra for these three time segments are shown in the middle panel of Figure~{\MyFigC}.
The robust low-energy and high-energy segments in the electron spectra are also marked with yellow and cyan shadow, respectively.
From this panel, one can find that the morphology of the electron spectra varies with time in a pulse,
especially the morphology of the low-energy segment.
In terms of this pulse,
the electron spectra in its low-energy regime
can be very different from the standard fast-cooling pattern and even a broken power-law function.
Besides, we also perform similar spectral analysis for four pulses in this burst, which are in [7.8, 11.2]~s (marked with red color), [15.6, 17.0]~s (marked with green color),
[29.7, 31.5]~s (marked with blue color), and [49.0, 52.4]~s (marked with gray color), respectively.
Please see the details in the inset of the right panel of Figure~{\MyFigC}.
The obtained electron spectra are shown in the right panel of this figure with the same color as that marking on the studied time period.
In terms of these pulses, the low-energy electron spectra can be also very different from the standard fast-cooling pattern and even a broken power-law function, e.g., the pulse marked with green color.

{\bf \emph{GRB~160509A Analysis }}
It is clear that GRB~180720B consists of multiple emission episodes.
In this paragraph, we would like to perform the spectral analysis for a burst with single contiguous and pulse-like structure, taking GRB~160509A as an example.
GRB~160509A is a long burst with redshift $z=1.17$ and detected by \emph{Fermi} and \emph{Swift} satellites.
The obtained NaI 0 light curve of the prompt emission is shown in the inset of Figure~{\MyFigD}, where the brightest NaI detector (NaI 0 and NaI 3) and BGO (BGO 0) detectors are used for our analyses.
Four different time periods are selected, [10-13.35]s, [13.35-14.65]s, [14.65-20]s, and [20-25]s,
which are marked with green, red, blue, and gray colors, respectively.
The obtained electron spectrum from spectral fitting for each time period
is shown with the same color in this figure and also reported in Table~{\MyTabA}.
Same as Figure~{\MyFigC}, the robust low-energy and high-energy segments in the electron spectra are also marked with yellow and cyan shadow, respectively.
One can find that the low-energy electron spectra are very different from the standard fast-cooling pattern.
The low-energy electron spectra in the time periods of [10-13.35]s, [13.35-14.65]s, and [14.65-20]s
are presented as a narrow bump rather than a power-law function.
The electron spectrum in the time period of [20-25]s is rather soft compared with other three electron spectra.
However, its low-energy segment is presented as a power-law function with index $\sim {-1.4}$ rather than $-2$.

At the end of this section, we present the reason why only the low-energy and high-energy segment in our obtained electron spectra are robust.
This is owing to that the synchrotron emission of the electrons at the lowest energy segment makes a negligible contribution to the total radiation spectrum.
The synchrotron radiation spectrum of an individual electron is $f_{\nu}\propto\nu^{1/3}$ for $\nu << \nu_{\rm c}$.
Thus the electron spectrum with power-law index being much larger than $-1/3$ would make a negligible effect on the radiation spectrum.
Therefore, the outline of the lowest-energy segment of the electron spectrum may can not be obtained by fitting the synchrotron radiation spectrum.
{\bf To differentiate the lowest-energy segment from the robust low-energy segment, here we propose another simpler but more general method.
Taking the spectral analysis in [7.19, 8.17]~s of GRB~180720B as an example,
we fix $\log y_1$ at two different values around its first best fit value (for example, $\log y_1=5{\ \rm and}\ -5$ in here) and perform twice independent fit again.
The electron spectra obtained from twice fit are shown as two blue dash lines around the electron spectrum of the first fit result.
The overlap region of these three spectra would be recognized as the robust low-energy segment.
Conversely, the divergence region would be recognized as the lowest-energy segment.
}

\section{Conclusions and Discussions}\label{Sec:Conclusion}
More and more evidences indicate that synchrotron radiation is a promising mechanism for the prompt emission of GRBs.
However, the electron spectrum for the prompt emission is diverse in numerical calculations or simulations.
In this paper, we propose a method to estimate the electron spectrum using an empirical function, which is a four-order \Bezier curve (low-energy regime) jointed with a linear function (high-energy regime) in log-log coordinate.
In the synchrotron radiation scenario with electron spectrum described by our empirical function, the following two works are studied in this paper.
(1) The electron spectrum corresponding to Band radiation spectrum is investigated.
We find that the exponential transition of Band radiation spectrum is more abrupt
compared with that of the synchrotron radiation spectrum based on a broken power-law electron spectrum.
Moreover, such exponential transition required a fine-tuned electron spectrum, which is hardly produced in real situation.
Then, we suggest that it may be inappropriate to use Band function to estimate the electron spectrum for the prompt emission of GRBs.
(2) We perform the spectral analysis on the observations of the prompt emission to estimate the electron spectrum.
GRB~180720B and GRB~160509A are studied as examples.
By performing spectral analysis for a series of time periods in these two bursts,
we find that the morphology of the electron spectrum in its low-energy regime evolves with time in a burst and even in a pulse.
In addition, it can be curved in some time periods,
which is very different from the standard fast-cooling pattern (i.e., $n\propto \gamma_{\rm e}^{-2}$) and even a power-law function.

{\bf Our proposed method is used to estimate the electron spectrum for the prompt emission,
without specifying a certain physical model for the electron spectrum.
In this paper, we focus on the synchrotron radiation scenario.
Actually, one could imagine convolving this electron spectrum with other emission kernels
may also get equally well-fitting solutions (pointed out by the referee).
It would be very interesting to investigate the shape of the electron spectrum
with other emission kernels.
}

\acknowledgments
We thank the anonymous referee of this work for useful comments and suggestions that improved the paper.
We also thank for Qi Wang and Zhi-Lin Chen for the useful discussions and suggestions.
This work is supported by
the National Natural Science Foundation of China
(grant Nos. 11773007, 11533003, U1938106, 11851304, U1731239),
the Guangxi Science Foundation
(grant Nos. 2018GXNSFFA281010, 2017AD22006, 2018GXNSFGA281007, 2018GXNSFDA281033),
and the Innovation Team and Outstanding Scholar Program in Guangxi Colleges.
We acknowledge the use of public data from the Fermi Science Support Center (FSSC).

\software{Xspec (\citealp{Arnaud-1996-ASPC.101.17A};
\citealp{Atwood_WB-2009-Abdo_AA-ApJ.697.1071A}),
gtBurst ({https://github.com/giacomov/gtburst}),
SciPy (\citealp{Jones-2001-Oliphant}), emcee (\citealp{Foreman-2013-Hogg-PASP.125.306F}), threeML (\citealp{Vianello_G-2015-arXiv})}

\clearpage
\begin{table}
\caption{Optimum value of parameters and the corresponding $\rm pgstat/d.o.f.$ in each time period.}
\label{MyTabA}
{\centering
\begin{tabular}{c|c|cccccccc}
\hline \hline
Burst & Time Period (s) & $\log y_1$ & $\log y_2$ & $\log y_3$ & $\log {y_4}$\tablenotemark{a} & $\log \gamma_{\rm m}$ & $p$ &
$B$\tablenotemark{a}& {\bf pgstat/d.o.f.} \\
\hline
 & $[7.17-8.19]$    &	 $0$   	 & 	$36.94$	& 	$40.86$	& 	$37.28$   & 	$4.34$ 	&	$-3.94$    &	$1473.89$		&	  $343.12/341$  \\
\cline{2-10}
 & $[8.19-8.70]$    &	 $0$   	 & 	$36.94$	& 	$40.86$	& 	$37.28$   & 	$4.34$ 	&	$-3.94$    &	$2000.09$	&	  $343.12/341$  \\
\cline{2-10}
 GRB & $[8.70-9.00]$    &	 $0$   	 & 	$36.94$	& 	$40.86$	& 	$37.28$   & 	$4.34$ 	&	$-3.94$    &	$1275.28$		&	  $343.12/341$  \\
\cline{2-10}
180720B & $[7.8-11.2]$     &      $0$   	 & 	$38.07$	& 	$37.23$ 	& 	$36.36$ 	& 	$4.02$ 	&	$-4.18$    &	$1863.36$	&	  $428.06/341$ \\
\cline{2-10}
 & $[15.6-17.0]$   &	 $0$   	 & 	$27.05$	& 	$38.60$ 	& 	$37.04$ 	& 	$4.04$ 	&	$-3.75$    &	$1347.31$		&	  $422.78/341$  \\
\cline{2-10}
 & $[29.7-31.5]$  & 	$22.77$    & 	$38.39$	& 	$37.24$ 	& 	$36.10$ 	& 	$3.93$ 	&	$-6.21$     &	$2603.91$	&	  $409.18/348$ \\
\cline{2-10}
 & $[49.0-52.4]$ & 	$20.56$    & 	$38.56$	& 	$37.62$	& 	$36.42$ 	& 	$4.00$ 	&	$-4.99$    &	$1248.34$	&	  $412.99/348$ \\
\hline
 & $[10.0-13.35]$ & 	$-30$    & 	$5.95$	& 	$41.83$	& 	$37.34$ 	& 	$3.82$ 	&	$-3.64$    &	$3099.20$	&	  $534.26/342$ \\
\cline{2-10}
GRB & $[13.35-14.65]$ & 	$-20$    & 	$10.21$	& 	$40.23$	& 	$36.65$ 	& 	$3.77$ 	&	$-4.08$    &	$3001.39$	&	  $342.40/341$ \\
\cline{2-10}
160905A & $[14.65-20.0]$ & 	$-10$    & 	 $14.69$	& 	$38.77$	& 	$36.33$ 	& 	$3.84$ 	&	$-3.70$    &	$1521.75$	&	  $561.99/312$ \\
\cline{2-10}
 & $[20.0-25.0]$ & 	$23.24$    & 	 $39.29$	& 	$37.69$	& 	$36.10$ 	& 	$3.90$ 	&	$-5.09$    &	$2521.53$	&	  $400.12/307$ \\
\hline \hline
\end{tabular}
\tablenotetext{a}{The value of the quantities are fixed in the fitting.}}
\end{table}

\clearpage
\begin{figure}
\begin{center}
\includegraphics[angle=0, scale=0.5, trim=60 70 30 0]{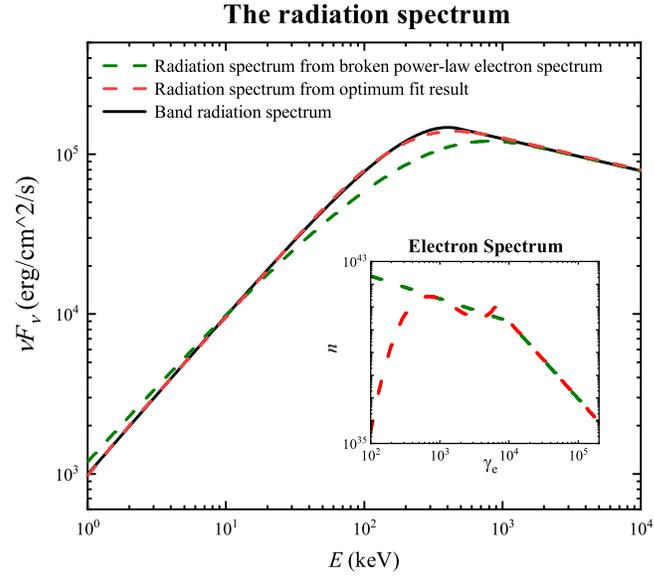}
\end{center}
\caption{Band radiation spectrum (black line) and the related electron spectra (inset panel).
Here a Band function with $\alpha=-1$, $\beta=-2.3$, and $E_{\rm p}=400$keV is discussed.
The inset plots the broken power-law electron spectrum and the electron spectrum obtained based on Equations~(\ref{Eq:Electrons_Spectrum}) and (\ref{EQ:SynPower}).
The corresponding synchrotron radiation spectra are shown with green and red dashed lines, respectively.
}\label{MyFigB}
\end{figure}
\clearpage
\begin{figure}
\begin{center}
\begin{tabular}{ccc}
\includegraphics[angle=0, scale=0.25, trim=60 70 60 0]{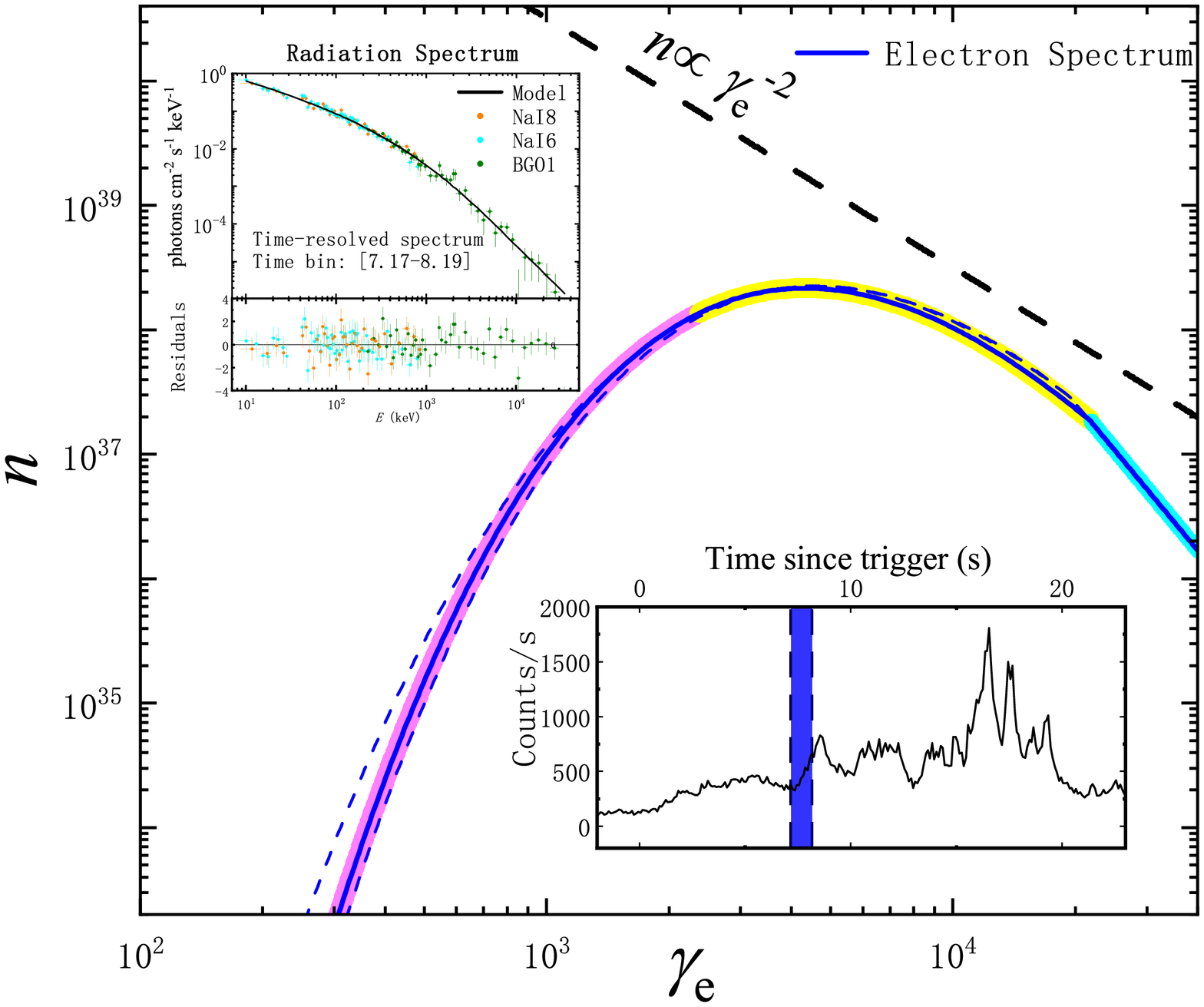}  &
\includegraphics[angle=0, scale=0.25, trim=60 50 60 0]{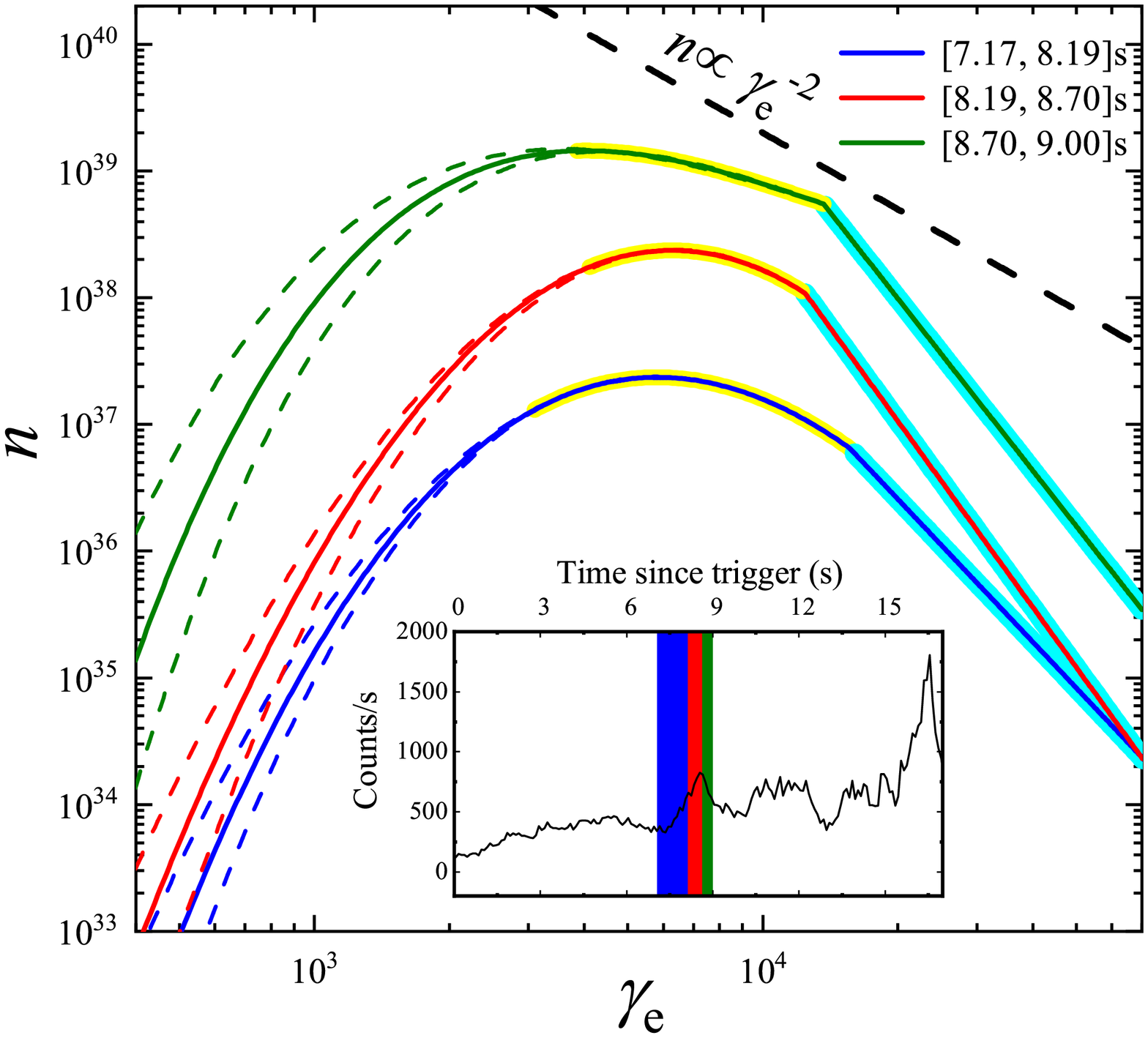}  &
\includegraphics[angle=0, scale=0.25 , trim=60 50 60 0]{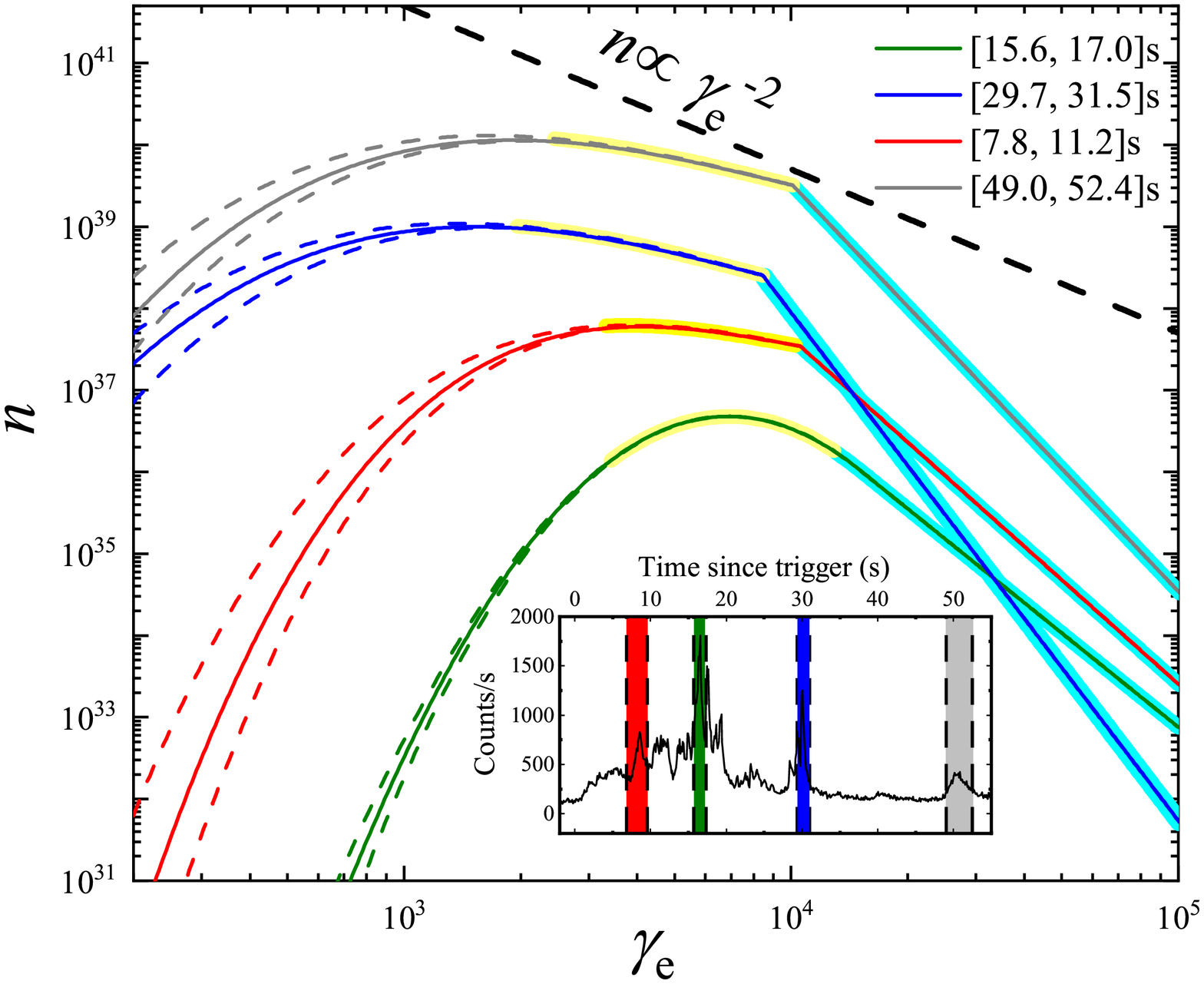}  \\
\end{tabular}
\end{center}
\caption{Electron spectra from our spectral fittings on GRB~180720B,
where the bottom inset in each panel shows the time periods (marked with different colors) for spectral fittings
and the corresponding electron spectrum is shown with solid lines and with the same color as that marking on the studied time period.
The dashed line below and above each solid lines are used to constrain the low-energy and high-energy segments
in our obtained electron spectrum.
In addition, the standard fast-cooling electron spectrum $n\propto \gamma_{\rm e}^{-2}$ is shown with black dashed line in each panel.
For convenient, the electron spectra of [8.19, 8.70]~s, [8.70, 9.00]~s, [7.8, 11.2]~s, [15.6, 17.0]~s, [29.7, 31.5]~s, and [49.0, 52.4]~s are shifted by timing 20, 30, 15, 0.1, 200, and 2000 factors, respectively.}
\label{MyFigC}
\end{figure}
\clearpage
\begin{figure}
\begin{center}
\begin{tabular}{c}
\includegraphics[angle=0, scale=0.75, trim=60 0 60 0]{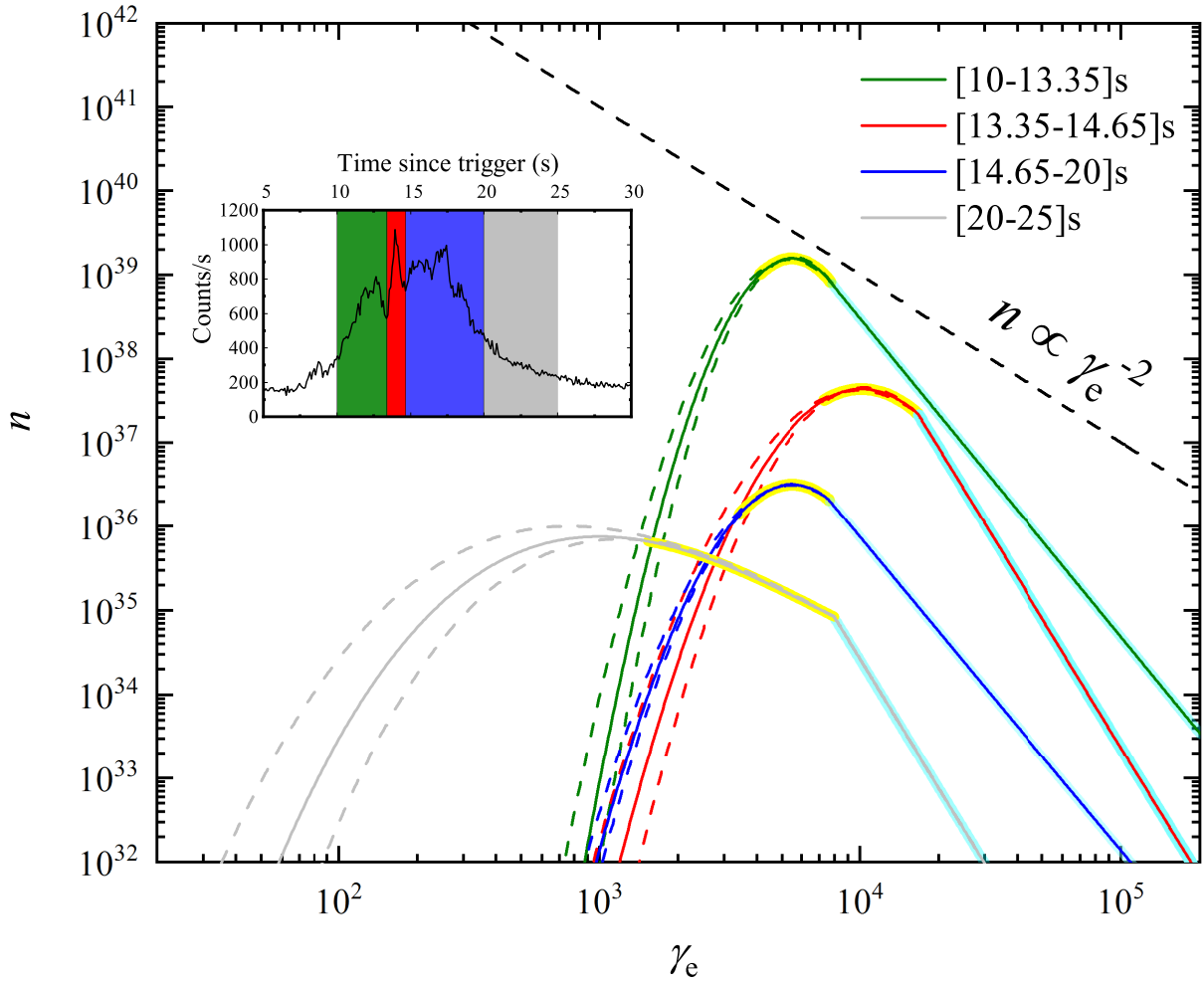}
\end{tabular}
\end{center}
\caption{Electron spectra from our spectral fittings on GRB~160509A and the upper inset shows the different time periods used for spectral analysis.
For convenient, the electron spectra of [13.35-14.65]~s, [14.65-20]~s, and [20-25]~s are also shifted by timing 0.01, 100, and 1/15, respectively.}
\label{MyFigD}
\end{figure}

\end{document}